\documentclass[10pt,conference,letterpaper]{IEEEtran}
\usepackage{epsfig,epsf,rotating,setspace,latexsym,amsmath,amssymb,amsfonts,bm,theorem,epstopdf,cite,authblk,bbm,mathrsfs,graphicx,caption, tabularx}
\usepackage{subcaption}
\usepackage{algorithm}
\usepackage[noend]{algpseudocode}
\usepackage{color}
\usepackage{mathtools}
\usepackage{multirow}
\usepackage{soul}
\usepackage[letterpaper, left=0.625in, right=0.625in, bottom=1.03in, top=0.72in]{geometry}
\algrenewcommand\algorithmicforall{\textbf{foreach}}
\algrenewcommand\algorithmicindent{.8em}

\newtheorem{theorem}{Theorem}
\newtheorem{lemma}{Lemma}
\newtheorem{corollary}{Corollary}

\newtheorem{definition}{Definition}

\newtheorem{example}{Example}
\newenvironment{Proof}[1]{\medskip\par\noindent{\bf Proof:\,}\,#1}{{\mbox{\,$\blacksquare$}\par}}

\newcommand{\cq}{{\mathcal{Q}}}
\newcommand{\cR}{{\mathcal{R}}}
\newcommand{\cw}{{\mathcal{W}}}
\newcommand{\cs}{{\mathcal{S}}}

\IEEEoverridecommandlockouts
\allowdisplaybreaks

\date{}
\title{Effect of Full Common Randomness Replication in Symmetric PIR on Graph-Based Replicated Systems}
\author{Shreya Meel \qquad Sennur Ulukus\\
\normalsize Department of Electrical and Computer Engineering\\
\normalsize University of Maryland, College Park, MD 20742 \\
\normalsize {\it smeel@umd.edu} \qquad {\it ulukus@umd.edu}}
 
\begin{document}

\maketitle

\begin{abstract}
We revisit the problem of symmetric private information retrieval (SPIR) in settings where the database replication is modeled by a simple graph. Here, each vertex corresponds to a server, and a message is replicated on two servers if and only if there is an edge between them. To satisfy the requirement of database privacy, we let all the servers share some common randomness, independent of the messages. We aim to quantify the improvement in SPIR capacity, i.e., the maximum ratio of the number of desired and downloaded symbols, compared to the setting with graph-replicated common randomness. Towards this, we develop an algorithm to convert a class of PIR schemes into the corresponding SPIR schemes, thereby establishing a capacity lower bound on graphs for which such schemes exist. This includes the class of path and cyclic graphs for which we derive capacity upper bounds that are tighter than the trivial bounds given by the respective PIR capacities. For the special case of path graph with three vertices, we identify the SPIR capacity to be $\frac{1}{2}$.
\end{abstract}

\section{Introduction}
Private information retrieval (PIR) \cite{chor}, is a cryptographic primitive where a user wishes to download their desired message in a database, potentially replicated in multiple non-colluding servers, without revealing the index of the message to any server. Over the past decade, PIR has been heavily studied from an information-theoretic perspective, with the primary metric of interest being the PIR capacity \cite{SJ17}, defined as the maximum number of message symbols retrieved per downloaded symbol. Beyond the basic setting  \cite{SJ17}, characterizing the PIR capacity under various configurations has been the focus of multiple works (see e.g., \cite{ banawan_pir_mdscoded, coded_colluding_2017, sun_eaves, kadhe_singleserver_pir} and \cite{ulukusPIRLC} for a  survey). One assumption in PIR is that, the privacy comes at the cost of the user learning a part of the messages they did not require. This is detrimental if the database contains sensitive information, and no database information beyond the desired message should be revealed to the user. To address this issue, symmetric PIR (SPIR) problem was formulated in \cite{spir_first}, introducing symmetric privacy requirements for both the user and the database.

To make SPIR feasible, the servers share some common randomness variable \cite{spir_first} prior to the communication protocol. The SPIR capacity and the minimum amount of randomness required to achieve the capacity was characterized in \cite{c_spir} in the basic fully-replicated database setting. This study was extended to other settings; e.g., SPIR with MDS coded messages \cite{skoglund_mds_spir, sun_spir_mds_mismatched}, SPIR with resilience against passive and active adversaries \cite{pir_spir_adversaries}, SPIR for multiple messages \cite{wang_mmspir}, SPIR with side information\cite{zhusheng_spir_pir}. All these works assume that, the entire database is available at all servers in a coded or an uncoded form. In practice, it may be judicious to constrain the replication of the databases at all servers, subject to cost and security reasons. Moreover, a database, though replicated, may be partially accessible to the user, subject to access control constraints  \cite{ali_dapac, meel_hetdapac}. Such scenarios motivate the study of PIR \cite{graphbased_pir, BU19, SGT23, meel_multi_pir} and SPIR \cite{meel2025symmetric} on databases whose replication pattern is described by a graph or hypergraph.

In the graph-replicated database model, each vertex corresponds to a server, and each edge represents a message stored on them. Different from the SPIR problem formulation in \cite{meel2025symmetric}, we do not restrict the availability of common randomness to the servers that share a message. Instead, we let the common randomness to be shared by all servers, independent of the graph modeling the database replication. Our goal in this paper is to study the improvement in capacity through full common randomness replication. For instance, we show that the capacity of a path graph on 3 vertices improves from $\frac{1}{3}$ in \cite{meel2025symmetric} to $\frac{1}{2}$ in our setting. In general, we characterize a capacity lower bound for all graphs whose PIR scheme admits a symmetric structure, followed by an upper bound for path and cyclic graphs. Similar to PIR capacity, our bounds on SPIR capacity depend on the explicit graph structure.

\section{System Model}
Consider a database of $K \geq 2$ independent messages given by the set $\mathcal{W}:=\{W_1,\ldots,W_K\}$. Each message comprises $L$ symbols selected uniformly at random from a finite field $\mathbb{F}_q$,
\begin{align}
    H(\cw)&=H(W_1)+\ldots+H(W_K)\\
        &=KL, \quad \text{in $q$-ary units.}
\end{align}
The messages are stored on $N\geq 2$ non-colluding servers, denoted by the set $\mathcal{S}=[N]$. Each message $W_k\in \cw$ is replicated exactly twice and stored on two distinct servers in $\cs$. We represent the database system by a simple, undirected graph $G=(V,E)$, with vertex set $V=\cs$, and edge set $E=\cw$. Hence, each vertex represents a server and each edge incident with a pair of servers, represents the message replicated at those servers. Further, we assume that $G$ is connected, i.e., there exists a path between each pair of vertices. 

\begin{figure}[t]
    \centering
    \includegraphics[width=0.45\textwidth]{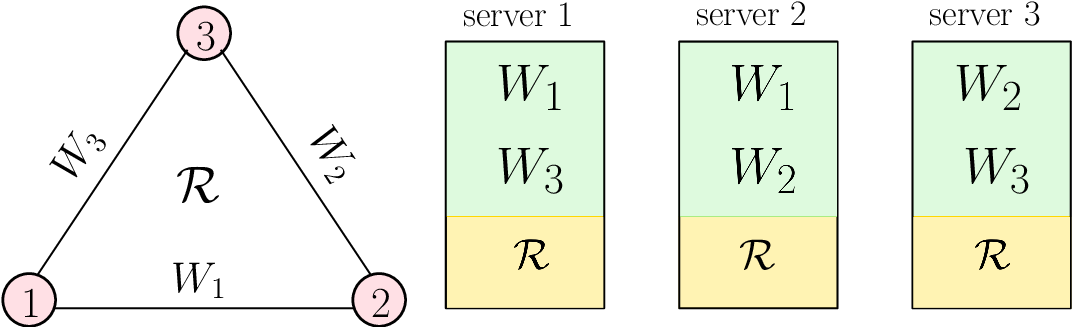}
    \caption{System model for $\mathbb{C}_3$.}
    \label{fig:sysmod_fr}
    \vspace{-0.4cm}
\end{figure}

In PIR, a user chooses an index $\theta \in [K]$, and wishes to privately retrieve the message $W_{\theta}\in \cw$, without revealing $\theta$ to any individual server. This is known as the user privacy. As an additional constraint in SPIR, the servers want to keep the messages other than $W_{\theta}$ private from the user. This is known as the database privacy. It is known \cite{spir_first} that SPIR is feasible only if the servers share some common randomness $\cR$. The symbols of $\cR$ are picked uniformly at random from $\mathbb{F}_q$, and are shared apriori by the servers, independent of $\cw$. The SPIR storage, modeled by the cyclic graph on three vertices is illustrated in Fig.~\ref{fig:sysmod_fr}. Note that, different from \cite{meel2025symmetric}, the common randomness $\cR$ is available to all servers, independent of the structure of $G$. 

Let $\cq$ represent the randomness in the schemes followed by the user to retrieve the messages. Since $\cq$ is decided prior to choosing the message index, it is independent of $\theta$. Further, $\theta$ and $\mathcal{Q}$ are independent of $\mathcal{W}$ and $\mathcal{R}$, since the user has no information of the content stored at the servers.

To retrieve $W_k$, the user generates $N$ queries $Q_1^{[k]}, \ldots, Q_N^{[k]}$ employing $\cq$, 
\begin{align}\label{eq:query_randomness}
    H(Q_1^{[k]},\ldots,Q_N^{[k]}|\cq) = 0,
\end{align}
and sends $Q_i^{[k]}$ to server $i$. Upon receiving the query, server $i$ responds with an answer $A_i^{[k]}$. Let $\mathcal{W}_i$ denote the set of messages stored at server $i$. Then, 
\begin{align}\label{eq:answer_deterministic}
    H(A_i^{[k]}|Q_i^{[k]}, \mathcal{W}_i, \mathcal{R}) = 0.
\end{align}
The answers and queries should satisfy the requirements of user privacy, reliability and database privacy, as defined next. For user privacy, the query and answer for each server should be identically distributed, irrespective of $\theta$, i.e., for every $i\in [N]$ and any index $k$,
\begin{align}\label{eq:user_privacy}
    (Q_i^{[k]}, A_i^{[k]},\mathcal{W}_i, \mathcal{R}) \sim (Q_i^{[1]}, A_i^{[1]},\mathcal{W}_i, \mathcal{R}).
\end{align}
For reliability, the user should be able to exactly recover their requested message $W_{k}$, using the answers from all the servers,
\begin{align}
    H(W_k|A_1^{[k]},\ldots,A_N^{[k]}, \cq) = 0.\label{eq:reliability}
\end{align}
For database privacy, no information on the desired message complement $W_{\overline{k}}:=\cw\setminus \{W_k\}$ should be revealed to the user, 
\begin{align}\label{eq:database_privacy2}
    H(W_{\bar{k}};A_1^{[k]},\ldots,A_N^{[k]},Q_1^{[k]},\ldots,Q_N^{[k]}|\cq)=0.
\end{align} 
If an SPIR scheme with fully replicated randomness, simultaneously satisfies \eqref{eq:user_privacy}, \eqref{eq:reliability}, and \eqref{eq:database_privacy2}, it is said to  be achievable. The rate $R_{\text{F-R}}$ of an SPIR scheme on $G$ is defined as the ratio between the number of desired message symbols and the total number of downloaded symbols,  
\begin{align}
    R_{\text{F-R}}(G)\triangleq\frac{L}{\sum_{i=1}^N H(A_i^{[k]})}.
\end{align} 
To distinguish this from the setting of \cite{meel2025symmetric}, we include the subscript F-R, to imply fully-replicated. The capacity $\mathscr{C}_{\text{F-R}}(G)$ is defined as the supremum over all achievable rates. To characterize the amount of common randomness for an SPIR scheme, we define the total randomness ratio,
\begin{align}
    \rho_{total}(G) \triangleq \frac{H(\cR)}{L},
\end{align}
as the size of total common randomness $\mathcal{R}$, required to be shared by all servers, relative to the message size. 

It is intuitive that, replicating the common randomness in all servers leads to better interference alignment, implying savings in the download cost. Therefore, for any graph $G$, the capacity $\mathscr{C}_{\text{F-R}}(G)$ is greater than $\mathscr{C}(G)$, and the inequality 
\begin{align}\label{relation_capacities}
   \mathscr{C}(G)<\mathscr{C}_{\text{F-R}}(G)<\mathscr{C}_{PIR}(G)
\end{align}
holds, where $\mathscr{C}_{PIR}(G)$ denotes the capacity of PIR on $G$, i.e., without the database privacy constraint. Therefore, $\mathscr{C}_{PIR}(G)$ is a trivial upper bound on $\mathscr{C}_{\text{F-R}}(G)$.
\section{Main Results}
Our first result characterizes a lower bound on $\mathscr{C}_{\text{F-R}}(G)$ by constructing an SPIR scheme from a PIR scheme on $G$. We consider the class of schemes wherein the PIR answer from every server is a set of $t$-sums (sum of $t$ symbols from distinct messages stored in the server) where $t\geq 1$. Additionally, the PIR scheme should satisfy the symmetric retrieval property \cite{meel_multi_pir}, as defined next.

\begin{definition}[Symmetric Retrieval Property (SRP)]
A graph-based PIR scheme is said to satisfy the symmetric retrieval property if, for any realization $k$ of $\theta$, the number of symbols of $W_k$ retrieved from each of the servers storing $W_k$, is equal. Equivalently,
$H(A_{i}^{[k]}|Q_{i}^{[k]},\cw_{i}\setminus\{W_k\})=H(A_{j}^{[k]}|Q_{j}^{[k]},\cw_{j}\setminus\{W_k\})=\frac{H(W_k)}{2}, $if $W_k$ is replicated on servers $i$ and $j$. 
\end{definition}

\begin{theorem}\label{thm:spir from pir scheme}
    Given a PIR scheme on $G$ with $N$ vertices and $K$ edges, that satisfies SRP, there exists an SPIR scheme with fully-replicated randomness on $G$. Let $L'$ be the number of symbols per message (note that SRP forces the PIR scheme to have even $L'$), and $D'$ be the total number of downloaded symbols of the PIR scheme, then the following SPIR rate,
    \begin{align}\label{eq:rate_G}
        R_{\text{F-R}}(G)=\frac{L'x}{D'x+Ny}\leq \mathscr{C}_{\text{F-R}}(G)
    \end{align}
    is achievable, where $x=\frac{\ell}{L'/2}$, $y=\frac{\ell}{N-1}$, $\ell=\text{lcm}(L'/2,N-1)$ and $\text{lcm}$ denotes the least common multiple. The corresponding randomness ratio is,
    \begin{align}\label{eq:rho_total_G}
        \rho_{total}(G)=\frac{K-1}{2}+\frac{Ny}{L'x}.
    \end{align}
\end{theorem}
The algorithm in Section~\ref{sec:achievable_algo} yields Theorem~\ref{thm:spir from pir scheme}. 

\begin{corollary}
    The capacity-achieving schemes of path \cite{our_journal2025} and cyclic \cite{BU19} graphs satisfy the SRP. In fact, if $G$ is a path graph $\mathbb{P}_N$ or a cyclic graph $\mathbb{C}_N$, \eqref{eq:rate_G} can be reduced to
    \begin{align}
        R_{\text{F-R}}(\mathbb{P}_N) & = \frac{2}{N+\frac{N}{N-1}},\label{rate_path}\\
        R_{\text{F-R}}(\mathbb{C}_N) & = \frac{2}{N+1+\frac{N}{N-1}}\label{rate_cyclic}.
    \end{align}
    This strictly improves upon $\mathscr{C}(\mathbb{P}_N)=\mathscr{C}(\mathbb{C}_N)=\frac{1}{N}$ in \cite{meel2025symmetric} and is related to the respective PIR capacities by
    \begin{align}
        \frac{1}{R_{\text{F-R}}(G)}=\frac{1}{\mathscr{C}_{PIR}(G)}+\frac{N}{2(N-1)}.
    \end{align}
    Further, \eqref{eq:rho_total_G} simplifies to $\rho_{total}(G)=
    {R_{\text{F-R}}(G)}^{-1}-1$.   
\end{corollary}

\begin{theorem}\label{thm:capacity of P_3}
    The SPIR capacity of $\mathbb{P}_3$ (see Fig.~\ref{fig:spir_p3_fullyrep}) is $\mathscr{C}_{\text{F-R}}(\mathbb{P}_3)=\frac{1}{2}$ with the optimal $\rho_{total}(\mathbb{P}_3)=1$. 
\end{theorem}
We prove this through an achievable scheme (Section~\ref{sec:P3 achievable}) and a matching upper bound (Section~\ref{sec:P3 converse}). Next, we generalize this upper bound to $\mathbb{P}_N$ and $\mathbb{C}_N$. Both these bounds improve upon the trivial bounds $\mathscr{C}_{PIR}(\mathbb{P}_N)$ and $\mathscr{C}_{PIR}(\mathbb{C}_N)$.

\begin{theorem}\label{thm:upperbnd_capacity_fr}
    The $\mathscr{C}_{\text{F-R}}(G)$ of  path and cyclic graphs satisfy
    \begin{align}
        \mathscr{C}_{\text{F-R}}(\mathbb{P}_N)&\leq \frac{2}{N+\frac{2}{N-1}}, \\
        \mathscr{C}_{\text{F-R}}(\mathbb{C}_N)&\leq \frac{2}{N+1+\frac{1}{N-1}}.
    \end{align}
\end{theorem}
The proof of Theorem~\ref{thm:upperbnd_capacity_fr} is provided in Section~\ref{sec:upper_bnd path and cycle}.

\section{SPIR from PIR Schemes}
In this section, we describe the algorithm for Theorem~\ref{thm:spir from pir scheme}, followed by the achievability proof of Theorem~\ref{thm:capacity of P_3}. Before the algorithm description, we present its idea with two examples.

\subsection{Motivating Examples}

\begin{example}[Path graph]\label{ex:path}
Consider the SPIR system for $\mathbb{P}_3$ as shown in Fig.~\ref{fig:spir_p3_fullyrep}. Suppose, $L=4$ and $H(\cR)=5$. Let the message and randomness symbols, after private and independent permutations by the user be denoted by $W_1=(a_1,\ldots,a_4)$, $W_2=(b_1,\ldots,b_4)$ and $\cR=(s_1,\ldots,s_5)$. The answers of the SPIR scheme are shown in Table~\ref{tab:spir_p3_fully_rep}. First, the user downloads a unique randomness symbol from each server. Next, the user performs two repetitions of the PIR scheme on $\mathbb{P}_3$, with the modification that each downloaded answer is mixed with randomness symbols from $\cR$. For instance, if $\theta=1$, we add a new randomness symbol $s_3,s_5$ to $b_2$ and $b_4$, respectively. In the first repetition, we add $s_1$ (downloaded from server 2) to $a_1$, and $s_2$ (downloaded from server 1) to $a_2$. In the second repetition, we add $s_4$ (downloaded from server 3) to $a_3$ and $a_4$. This gives $R_{\text{F-R}}(\mathbb{P}_3)=\frac{4}{9}=\frac{2}{3+\frac{3}{2}}$ and $\rho_{total}(\mathbb{P}_N)=\frac{5}{4}=\frac{9}{4}-1$.

\begin{figure}[t]
    \centering
    \includegraphics[width=0.4\textwidth]{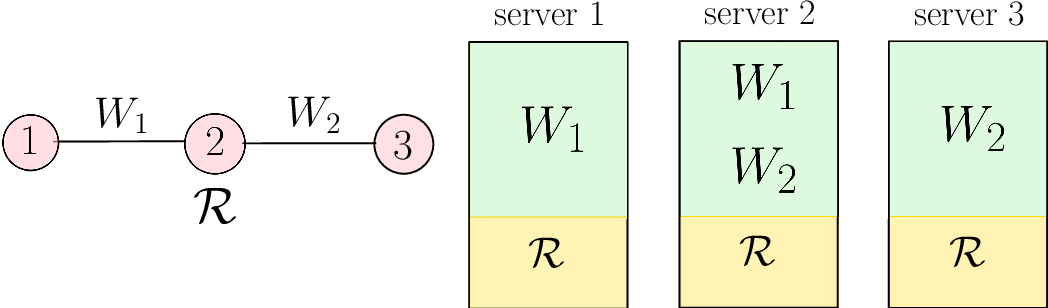}
    \caption{SPIR model for $\mathbb{P}_3$.}
    \label{fig:spir_p3_fullyrep}
\end{figure}

\begin{table}[h!]
\centering
\begin{tabular}{|c|c|c|c|}
\hline
$\theta=1$ &database 1 & database 2& database 3\\
\hline
&$s_2$ & $s_1$ & $s_4$\\
rep. 1&$a_1+s_1$ & $a_2+b_2+s_2+s_3$ & $b_2+s_3$\\
rep. 2&$a_3+s_4$ & $a_4+b_4+s_4+s_5$ & $b_4+s_5$ \\
\hline
\hline
$\theta=2$ &database 1 & database 2& database 3\\
\hline
&$s_5$ & $s_3$ & $s_1$\\
rep. 1&$a_1+s_1$ & $a_1+b_1+s_1+s_2$ & $b_2+s_3$\\
rep. 2&$a_3+s_4$ & $a_3+b_3+s_4+s_5$ & $b_4+s_5$ \\
\hline
\end{tabular}
\caption{Answer table for $\mathbb{P}_3$.}
\label{tab:spir_p3_fully_rep}
\end{table}
\end{example}

\begin{example}[Cyclic graph]
For the SPIR scheme on $\mathbb{C}_N$ with $N=3$, we build upon the scheme in \cite{BU19}. The answer table when $W_1$ is desired is shown in Table~\ref{tab:spir_c3_fully_rep}. Corresponding to each undesired symbol, we assign a new distinct randomness symbol to be added to the answer. For each desired symbol downloaded from server 1 (server 2), a randomness symbol downloaded from server 2 (server 1) and server 3 is assigned. This gives $R_{\text{F-R}}(\mathbb{C}_N)=\frac{4}{11}=\frac{2}{3+1+\frac{3}{2}}$ and $\rho_{total}(\mathbb{C}_N)=\frac{7}{4}=\frac{11}{4}-1$.

\begin{table*}[h!]
\centering
\begin{tabular}{|c|c|c|c|}
\hline
&database 1 & database 2& database 3\\
\hline
& $s_7,s_9,s_{11}$ & $s_1,s_3,s_5$ & $s_{13},s_{15},s_{17}$\\
\hline
\parbox[t]{2mm}{\multirow{3}{*}{\rotatebox[origin=c]{90}{rep. 1}}} &$a_1+s_1,c_1+s_2$ & $a_4+s_7, b_1+s_8$ & $b_2+s_{10}, c_2+s_4$\\
& $a_2+c_2+s_3+s_4$ & $a_5+b_2+s_9+s_{10}$ & $b_1+c_3+s_8+s_6$\\
& $a_3+c_3+s_5+s_6$ & $a_6+b_3+s_{11}+s_{12}$ & $b_3+c_1+s_{12}+s_2$\\
\hline
\parbox[t]{2mm}{\multirow{3}{*}{\rotatebox[origin=c]{90}{rep. 2}}}  &$a_7+s_{13},c_7+s_{14}$ & $a_{10}+s_{13}, b_7+s_{19}$ & $b_8+s_{20}, c_8+s_{16}$\\
& $a_8+c_8+s_{15}+s_{16}$ & $a_{11}+b_8+s_{15}+s_{20}$ & $b_7+c_9+s_{19}+s_{18}$\\
& $a_9+c_9+s_{17}+s_{18}$ & $a_{12}+b_9+s_{17}+s_{21}$ & $b_9+c_7+s_{14}+s_{21}$\\
\hline
\end{tabular}
\caption{Answer table for $\mathbb{C}_3$ when $\theta=1$.}
\label{tab:spir_c3_fully_rep}
\end{table*}
\end{example}

\subsection{Algorithm Description}\label{sec:achievable_algo}
It is clear from the examples that our scheme exhibits two key properties. 1) The user downloads an equal number $y$ of uncoded randomness symbols from every server, and 2) if the randomness symbols are removed from the answers, our scheme reduces to $x$ repetitions of the starting PIR scheme.

Let $W_k$ be the desired message, replicated on servers $i, j\in [N]$. To determine $x$ and $y$, for a given PIR scheme, observe that, in each repetition $x$, the user retrieves $xL'/2$ symbols of $W_k$, masked by a symbol of $\cR$, from each server $i$ and $j$. For decoding the $xL'/2$ symbols of $W_k$ retrieved from each of server $i$ and $j$, the user downloads the corresponding $xL'/2$ randomness symbols from the remaining servers, i.e., $[N]\setminus \{i\}$ and $[N]\setminus \{j\}$, respectively. Therefore, by property 1, this implies, $x$ and $y$ are least positive integers satisfying
\begin{align}\label{eq:relation of x and y}
    x\frac{L'}{2}=(N-1)y.
\end{align}
Clearly, \eqref{eq:relation of x and y} is equal to $\text{lcm}(L'/2,N-1)$ by the definition of least common multiple. Now, we describe our algorithm steps.
\paragraph{Index permutation} For the SPIR scheme, let $L=xL'$ and $H(\cR)=Ny+\left(\frac{K-1}{2}\right)L$. The user applies private and independent permutations to the symbols of the $K$ messages, with $W_\ell=\left(w_\ell(1), \ldots, w_\ell(L)\right)$ and $\cR = (s_1,\ldots,s_{Ny+L(K-1)/2})$.
\paragraph{Query generation} The user follows the query structure of the starting PIR scheme. Queries sent to the servers correspond to $x$ repetitions of the scheme, where in each repetition, the user queries for $L'$ new symbols of $W_{k}$. Corresponding to each queried symbol, the user sends the following common randomness assignment to each server. 
\paragraph{Common randomness assignment} The user assigns a unique randomness symbol to each queried message symbol $W_{\ell}\in W_{\overline{k}}$ across all repetitions. Note that, because of SRP, to maintain user privacy, we query $L'/2$ symbols of $W_\ell$ from every server in the PIR scheme. For reliability of the PIR scheme through interference cancellation, we require the same $L'/2$ symbols to be queried from both servers where $W_\ell$ is replicated. This assigns the $(K-1)L/2$ symbols of $\cR$. To the first $y=\frac{L}{2(N-1)}$ symbols of $W_k$ queried from server $i$ and $j$, the user assigns distinct randomness symbols $s_{j_1},\ldots,s_{j_y}$ and $s_{i_1},\ldots,s_{i_y}$, respectively. To each of the remaining $L/2-y=(N-2)y$ queried symbols of $W_k$ from each of server $i$ and $j$, the user assigns the same randomness symbol from the remaining set $\{s_{n_1},\ldots,s_{n_y}, n\in [N]\setminus\{i,j\}\}$ of $\cR$.  
\paragraph{Answer formation} The user downloads the randomness symbols $s_{n_1},\ldots,s_{n_y}$ from server $n\in [N]$, followed by $x$ repetitions of PIR answers. To each answer, the servers add randomness symbols, with each message symbol in the answer added to the assigned randomness symbol. That is, a $t$-sum in the answer is padded with $t$ symbols from $\cR$.

This completes the algorithm, with $R_\text{F-R}$ and $\rho_{total}$ derived by direct computation. Next, we will show that our algorithm outputs a feasible SPIR scheme.

\textit{Reliability:} By reliability of the PIR scheme, in repetition $u\in [x]$, the user retrieves $w_k((u-1)L'+1),\ldots,w_k(u L')$ added to an already downloaded symbol of $\cR$. Using this, they recover $L=xL'$ symbols of $W_k$. Specifically, $s_{j_1},\ldots,s_{j_y}$ and $s_{i_1},\ldots,s_{i_y}$ cancel out the randomness symbols to retrieve the first $y$ symbols of $W_k$ from server $i$ and $j$, respectively. Further, $s_{n_1},\ldots,s_{n_y}$ downloaded from servers $n\in[N]\setminus \{i,j\}$ are used as common side information to retrieve the remaining $L-2y$ symbols of $W_k$ from both servers $i$ and $j$. 

\textit{User privacy:} From the perspective of each server, the queried message symbols are independent of $\theta$, due to the PIR scheme. Further, the randomness symbols directly downloaded by the user, and those added to the PIR answers, appear to be uniformly chosen from $\cR$, irrespective of $\theta$, since the private permutation on $\cR$ hides the common randomness assignment.

\textit{Database privacy:} No information on $W_{\overline{k}}$ is revealed to the user, since every non-desired message symbol is protected by a unique symbol of $\cR$, which is distinct from those downloaded by the user in the scheme. 

\subsection{Achievability for $\mathbb{P}_3$}\label{sec:P3 achievable}
We demonstrate a simple scheme on $\mathbb{P}_3$ that achieves a higher rate of $R_{\text{F-R}}(\mathbb{P}_3)=\frac{1}{2}$ than $\frac{4}{9}$ in Example~\ref{ex:path} with $\rho_{total}(\mathbb{P}_3)=1<\frac{5}{4}$. With $L=L'=2$, the answers are shown in Table~\ref{tab:capacity_p3_scheme}, which in the absence of $s_1,s_2$, reduce to the PIR scheme on $\mathbb{P}_3$. The improvement comes from adding a single randomness symbol in the answer of server 2, and downloading a randomness symbol from server 2 alone. Notice that $\mathbb{P}_3$ is identical to a star graph, which enables this, unlike $\mathbb{P}_N, N\geq 4$. 
\begin{table}[h!]
\centering
\begin{tabular}{|c|c|c|c|}
\hline
&database 1 & database 2& database 3\\
\hline
$\theta=1$ &$a_1+s_1$ & $s_1,a_2+b_2+s_2$ & $b_2+s_2$\\
\hline
$\theta=2$ & $a_1+s_1$ & $s_2, a_1+b_1+s_1$ & $b_2+s_2$\\
\hline
\end{tabular}
\caption{SPIR answer table for $\mathbb{P}_3$ for the improved scheme.}
\label{tab:capacity_p3_scheme}
\end{table}
\section{Upper Bound on SPIR Capacity}
We have that any feasible SPIR scheme with fully-replicated randomness, $\cR$ satisfies
\begin{align}
    H(\cR)=&H(\cR|\cw,\cq)+H(A^{[k]}_{1:N}|\cw,\cR,\cq)\\
    =& H(A_{1:N}^{[k]},\cR|\cw,\cq)\\
    \geq & H(A_{1:N}^{[k]}|\cw,\cq)\\
    =& H(A_{1:N}^{[k]}|\cq) -I(\cw;A_{1:N}^{[k]}|\cq)\\
    \stackrel{\eqref{eq:reliability}\eqref{eq:database_privacy2}}{=} &H(A_{1:N}^{[k]}|\cq)-L. \label{eq:lowerbnd_randomness}
\end{align}
For the capacity upper bounds, we state the following lemmas. The first lemma generalizes \cite[Lemma 2]{c_spir}.
\begin{lemma}\label{lem:answer n depends only on query n}
    For any $\mathcal{J}\subseteq [K]$, 
    \begin{align}
      H(A_n^{[k]}|W_{\mathcal{J}},\cR, Q_n^{[k]})=H(A_n^{[k]}|W_{\mathcal{J}},\cR, Q_n^{[k]},\cq).
    \end{align}
\end{lemma}
The next lemma is a consequence of user privacy \eqref{eq:user_privacy} and graph-replicated databases. 
\begin{lemma}\label{lem: conseq of user priv}
    For any $\mathcal{J}\subseteq [K]$, and any server $n\in [N]$, for all $k, k'\in [K]$, we have
\begin{align}
    H(A_n^{[k]}|W_\mathcal{J},\cR,Q_n^{[k]}) &= H(A_n^{[k']}|W_\mathcal{J},\cR,Q_n^{[k']}),\\
        H(A_n^{[k]}|W_\mathcal{J},Q_n^{[k]}) &= H(A_n^{[k']}|W_\mathcal{J},Q_n^{[k']}).
\end{align}
\end{lemma}
The following lemma is an extension of \cite[Lemma 3]{SGT23}.

\begin{lemma}\label{sum of answer entropies}
For $k\in [K]$, let $W_k$ be replicated on servers servers $i$ and $j$. Then, 
\begin{align}
H(A_i^{[k]}|W_{\overline{k}},\cR,\cq)+H(A_j^{[k]}|W_{\overline{k}},\cR,\cq)\geq L.
\end{align}
\end{lemma}

\subsection{Converse for $\mathbb{P}_3$}\label{sec:P3 converse}
Suppose $W_1$ is the desired message. Then,
\begin{align}
    L \stackrel{\eqref{eq:reliability}}{=}&H(W_1|\cq)-H(W_1|A_{1:3}^{[1]},\cq)\label{eq:starting eqn p3}=I(A^{[1]}_{1:3};W_1|\cq)\\
    \stackrel{\eqref{eq:query_randomness}}{=}&H(A_{1:3}^{[1]}|\cq)-H(A_1^{[1]}|W_1,Q_1^{[1]},\cq)\notag \\
    &-H(A_{2:3}^{[1]}|W_1,A^{[1]}_1,\cq) \label{eq:first step generalize}\\
    \leq&H(A_{1:3}^{[1]}|\cq)-H(A_1^{[1]}|W_1,Q_1^{[1]})\notag\\
    &-H(A_{2:3}^{[1]}|W_1,A^{[1]}_1,\cq,\cR)\label{eq:follows from lemma 2}\\
    =& H(A_{1:3}^{[1]}|\cq)-H(A_1^{[2]}|W_1,Q_1^{[2]})-H(A_2^{[1]}|W_1,\cq,\cR)\notag \\
    &-H(A_3^{[1]}|W_1,A_2^{[1]},\cq,\cR) \label{eq:consequence of lemma 1}\\
    \stackrel{\eqref{eq:database_privacy2}\eqref{eq:user_privacy}}{=} & H(A_{1:3}^{[1]}|\cq)-H(A_1^{[1]}|Q_1^{[1]})-H(A_2^{[1]}|W_1,\cq,\cR)\notag \\
    &-H(A_3^{[1]}|W_1, A_2^{[1]},\cq,\cR) \\
   \leq& H(A_{1:3}^{[1]}|\cq)-H(A_1^{[1]}|\cq)-H(A_2^{[1]}|W_1,\cq,\cR)\notag\\
    &-H(A_3^{[1]}|W_1,W_2,A_2^{[1]},\cq,\cR)\\
    \stackrel{\eqref{eq:answer_deterministic}}{=}& H(A_{1:3}^{[1]}|\cq)-H(A_1^{[1]}|\cq)-H(A_2^{[2]}|W_1,\cq,\cR) ,\label{eq:sum one way p3}
\end{align}
where \eqref{eq:follows from lemma 2} follows from Lemma~\ref{lem:answer n depends only on query n} and conditioning with $\cR$,  \eqref{eq:consequence of lemma 1} holds since $A_1^{[1]}$ is completely determined by $(W_1,\cq,\cR)$, and Lemma~\ref{lem: conseq of user priv}, and \eqref{eq:sum one way p3} also follows from Lemma~\ref{lem: conseq of user priv}. By interchanging the positions of $A_2^{[1]}$ and $A_3^{[1]}$ in \eqref{eq:consequence of lemma 1}, we obtain
\begin{align}
    L\leq H(A_{1:3}^{[1]}|\cq)-H(A_1^{[1]}|\cq)-H(A_3^{[2]}|W_1,\cq,\cR) .\label{eq:sum other way p3}
\end{align}
From \eqref{eq:sum one way p3} and \eqref{eq:sum other way p3}, and applying Lemma~\ref{sum of answer entropies}, we get
\begin{align}
L\leq& H(A_{1:3}^{[1]}|\cq)-H(A_1^{[1]}|\cq)\notag \\
&-\frac{1}{2}\left(H(A_2^{[2]}|W_1,\cq,\cR)+H(A_3^{[2]}|W_1,\cq,\cR)\right) \\
\leq & H(A_{1:3}^{[1]}|\cq)-H(A_1^{[1]}|\cq)-\frac{L}{2},\label{eq:answer except 1 p3}
\end{align}
which, since $H(A_{1:3}^{[1]}|\cq)-H(A_1^{[1]}|\cq)\leq H(A_2^{[1]}|\cq)+H(A_3^{[1]}|\cq)$, upon rearrangement gives
\begin{align}\label{eq:theta is 1}
    H(A_2^{[1]}|\cq)+H(A_3^{[1]}|\cq)\geq \frac{3L}{2}.
\end{align}
Repeating the steps in \eqref{eq:starting eqn p3}-\eqref{eq:answer except 1 p3} for $W_2$, we get
\begin{align}\label{eq:theta=2}
    H(A_1^{[2]}|\cq)+H(A_2^{[2]}|\cq)\geq \frac{3L}{2}.
\end{align}
Moreover, for $k=1,2$, by bounding $H(A_{1:3}^{[k]}|W_k,\cq)$ as
\begin{align}
   H(&A_{1:3}^{[k]}|W_k,\cq) \notag\\
   &= H(A_2^{[k]}|W_k,\cq)+H(A_1^{[k]},A_3^{[k]}|W_k,A_2^{[k]},\cq)\\
   &\geq H(A_2^{[k]}|\cq) + H(A_1^{[k]},A_3^{[k]}|W_1,W_2,\cR,A_2^{[k]},\cq)\\
   &= H(A_2^{[k]}|\cq) + H(A_1^{[k]},A_3^{[k]}|W_1,W_2,\cR,\cq)\\
   &=  H(A_2^{[k]}|\cq),
\end{align}
we get
\begin{align}
    H(A_1^{[k]}|\cq)+H(A_3^{[k]}|\cq)\geq L.\label{eq:sum entropy edges p3}
\end{align}

Finally,  summing \eqref{eq:sum one way p3}, \eqref{eq:sum other way p3} and \eqref{eq:sum entropy edges p3}, for $k=1,2$ by \eqref{eq:user_privacy}, we obtain
\begin{align}
2\big(H(A_1^{[k]}|\cq)+H(A_2^{[k]}|\cq)+H(A_3^{[k]}|\cq)\big)\geq 4L,\label{eq: p3 download bound}
\end{align}
which gives $\mathscr{C}(\mathbb{P}_3)\leq \frac{1}{2}$. To determine the bound on $H(\cR)$, from \eqref{eq:lowerbnd_randomness}, we have $H(\cR)\geq H(A_{1:3}^{[k]}|\cq)-L$, which satisfies 
\begin{align}
   H(A_{1:3}^{[1]}|\cq)-L&\geq H(A_1^{[k]}|\cq)+\frac{L}{2},\\
   H(A_{1:3}^{[k]}|\cq)-L&\geq H(A_2^{[k]}|\cq),\\
   H(A_{1:3}^{[2]}|\cq)-L&\geq H(A_3^{[k]}|\cq)+\frac{L}{2}.
\end{align}
From this, we obtain $3H(\cR)\geq H(A_1^{[k]}|\cq)+H(A_2^{[k]}|\cq)+H(A_3^{[k]}|\cq)+L\geq 3L$, i.e., $\rho_{total}(\mathbb{P}_3)\geq 1$.
\subsection{Proof of Upper Bounds for $\mathbb{P}_N$ and $\mathbb{C}_N$}\label{sec:upper_bnd path and cycle}
Recall that for $\mathbb{P}_N$, $K=N-1$, and $W_n$ is replicated on servers $n$ and $n+1$, for all $n\in [N-1]$ while for $\mathbb{C}_N$, $\cw_n = \{W_{n-1},W_{n}\} , n\in [N]$ with $W_0=W_N$. The next two lemmas lower bound the size of $A_{[1:N]\setminus\{n\}}^{[k]}$, conditioned on $A_n^{[k]}, W_k$ and $\cq$, for $\mathbb{P}_N$ and $\mathbb{C}_N$, respectively. We omit the proof of Lemma~\ref{lem:answer bound cycle} since it follows similarly as that of Lemma~\ref{lem:ans bound path}.
\begin{lemma}\label{lem:ans bound path}
For any feasible SPIR scheme on $\mathbb{P}_N$, we have
        \begin{align}
        H(A_{2:N}^{[1]}|A_1^{[1]},W_1,\cq)&\geq \frac{(N-2)L}{2}, \label{eq:answer of server 1}\\
            H(A_{1:N-1}^{[N-1]}|A_{N}^{[N-1]},W_{N-1},\cq) & \geq \frac{(N-2)L}{2}, \label{eq:answer of server N}\\
               H(A_{1:N\setminus \{n\}}^{[k]}|A_n^{[k]},W_k,\cq)&\geq \frac{(N-3)L}{2}, \notag\\ 
               k\in \{n-1,n\}, n\in &\{2,\ldots,N-2\}. \label{eq:answers of non-edge servers}
        \end{align}
\end{lemma} 
\begin{Proof}
To show \eqref{eq:answer of server 1}, we have that $H(A_{2:N}^{[1]}|A_1^{[1]},W_1,\cq)\geq H(A_{2:N}^{[1]}|A_1^{[1]},W_1,\cR,\cq)=H(A_{2:N}^{[1]}|W_1,\cR,\cq)$ where the equality is because $H(A_1^{[k]}|W_1,\cR,\cq)=0$. Now,
\begin{align}
      H(A^{[1]}_{2:N}&|W_1,\cR,\cq)\notag \\
      &= \sum_{n\in [2:N]}H(A_n^{[1]}|A^{[1]}_{2:n-1},W_1,\cR,\cq)\label{eq:answer bound}\\
      &\geq \sum_{n\in [2:N]} H(A_n^{[1]}|A^{[1]}_{2:n-1},W_{1:n-1},\cR,\cq)\\
      &= \sum_{n\in [2:N-1]} H(A_n^{[n]}|W_{1:n-1},\cR,\cq)\label{eq:Nth answer is 0 and Lemma 1}\\
      &\geq \sum_{n\in [2:N-1]} H(A_n^{[n]}|W_{\overline{n}},\cR,\cq)\label{eq:answer bound one way}.
  \end{align}
Here \eqref{eq:Nth answer is 0 and Lemma 1} holds since $H(A_N^{[1]}|W_{1:N-1},\cR,\cq)=0$ and by Lemma~\ref{lem: conseq of user priv}. 
Expanding \eqref{eq:answer bound} in reverse order of server indices,
  \begin{align}
      H&(A^{[1]}_{2:N}|A_1^{[1]},W_1,\cR,\cq)\notag \\
      &= \sum_{n\in [2:N]} H(A_{n}^{[1]}|A^{[1]}_{n+1:N},W_1,\cR,\cq)\\
      &\geq \sum_{n\in [2:N]} H(A_{n}^{[1]}|A^{[1]}_{n+1:N}, W_1, W_{n:N-1},\cR,\cq)\\
      &= \sum_{n\in [3:N]} H(A_{n}^{[n-1]}| W_1, W_{n:N-1},\cR,\cq)\\
      &\geq\sum_{n\in [3:N]}H(A_{n}^{[n-1]}|W_{\overline{n-1}},\cR,\cq).\label{eq:answer bound reversed}     
  \end{align}  
Summing \eqref{eq:answer bound one way} and \eqref{eq:answer bound reversed}, and using Lemma~\ref{sum of answer entropies}, we obtain
\begin{align}
   2 &H(A^{[1]}_{2:N}|A_1^{[1]},W_1,\cR,\cq) \notag \\
   & \geq \sum_{n=2}^{N-1} H(A_n^{[n]}|W_{\overline{n}},\cR,\cq)+H(A_{n+1}^{[n]}|W_{\overline{n}},\cR,\cq)\\
    &\geq  (N-2)L.
\end{align}
Proving \eqref{eq:answer of server 1} and \eqref{eq:answer of server N} follows similarly. For \eqref{eq:answers of non-edge servers}, proceeding similarly, and using $ H(A_{1:N\setminus \{n\}}^{[k]}|A_n^{[k]},W_k,\cR,\cq)\geq  H(A_{1:N\setminus \{n\}}^{[k]}|W_{n-1},W_n,\cR,\cq)$, $k=n-1,n$ yields the desired result.
\end{Proof}

To prove the upper bound for $\mathbb{P}_N$, for $k\in [N-1]$ we have 
\begin{align}
    \! \! L&=I(A_{1:N}^{[k]};W_k|\cq)=H(A_{1:N}^{[k]}|\cq)-H(A_{1:N}^{[k]}|W_k,\cq) \label{eq:routine step start}\\
    &=H(A_{1:N}^{[k]}|\cq)-H(A_n^{[k]}|\cq)-H(A_{1:N\setminus \{n\}}^{[k]}|A_n^{[k]},W_k,\cq)\label{eq: applying lemma user privacy}\\
    &\leq \sum_{n'\in [N]\setminus\{n\}}H(A_{n'}^{[k]}|\cq)-H(A_{1:N\setminus \{n\}}^{[k]}|A_n^{[k]},W_k,\cq).\label{eq:routine step end}
\end{align}
Bounding the second term by Lemma~\ref{lem:ans bound path} we obtain for all $k$,
\begin{align}
\sum_{n'\in [N]\setminus \{n\}} H(A_{n'}^{[k]}|\cq)\geq \begin{cases}
    \frac{NL}{2}, &n=1,N,\\
    \frac{(N-1)L}{2}, &n\in [2:N-1],
\end{cases}
\end{align}
by user privacy \eqref{eq:user_privacy}. Summing over all $n\in [N]$, we get
\begin{align}
     \sum_{n=1}^N &\sum_{n'\in [N]\setminus\{n\}} H(A_{n'}^{[k]}|\cq)\notag \\ &\geq 2\frac{NL}{2}+(N-2)\frac{(N-1)L}{2} = \frac{(N^2-N+2)L}{2},
\end{align}
resulting in
\begin{align}
     &(N-1)\sum_{n=1}^N H(A_n^{[k]}|\cq) \geq  \frac{(N^2-N+2)L}{2},
\end{align}
which gives 
$\frac{L}{\sum_{n=1}^N H(A_n^{[k]})}\leq\frac{L}{\sum_{n=1}^N H(A_n^{[k]}|\cq)}\leq \frac{2(N-1)}{N^2-N+2}=\frac{2}{N+\frac{2}{N-1}}$, proving the upper bound for $\mathbb{P}_N$. 

For $\mathbb{C}_N$, we need the following lemma, where $k=0$ implies $k=N$.
\begin{lemma}\label{lem:answer bound cycle}
 For any feasible SPIR scheme on $\mathbb{C}_N$, and $n\in [N]$,
    \begin{align}
        H(A_{1:N\setminus \{n\}}^{[k]}|A_n^{[k]},W_k,\cq)&\geq \frac{(N-2)L}{2}, \, k=n-1,n.
    \end{align}
\end{lemma}
Then, \eqref{eq:routine step start}-\eqref{eq:routine step end} and Lemma~\ref{lem:answer bound cycle} result in 
\begin{align}
        &(N-1)\sum_{n=1}^N H(A_n^{[k]}|\cq) \geq N\left(L+\frac{(N-2)L}{2}\right),
\end{align}
for any achievable scheme, which yields
\begin{align}
      &\frac{L}{\sum_{n=1}^N H(A_n^{[k]})}\leq \frac{2(N-1)}{N^2}=\frac{2}{N+1+\frac{1}{N-1}}. 
\end{align}
This completes the proof of Theorem~\ref{thm:upperbnd_capacity_fr}.
\bibliographystyle{unsrt}
\bibliography{references}

\begin{thebibliography}{10}

\bibitem{chor}
B.~Chor, E.~Kushilevitz, O.~Goldreich, and M.~Sudan.
\newblock Private information retrieval.
\newblock {\em Journal of the ACM}, 45(6):965--981, November 1998.

\bibitem{SJ17}
H.~Sun and S.~A. Jafar.
\newblock The capacity of private information retrieval.
\newblock {\em IEEE Trans. Inf. Theory}, 63(7):4075--4088, March 2017.

\bibitem{banawan_pir_mdscoded}
K.~Banawan and S.~Ulukus.
\newblock The capacity of private information retrieval from coded databases.
\newblock {\em IEEE Trans. Inf. Theory}, 64(3):1945--1956, January 2018.

\bibitem{coded_colluding_2017}
R.~Freij-Hollanti, O.~W. Gnilke, C.~Hollanti, and D.~A. Karpuk.
\newblock Private information retrieval from coded databases with colluding servers.
\newblock {\em SIAM J. Appl. Algebra and Geometry}, 1(1):647--664, 2017.

\bibitem{sun_eaves}
Q.~Wang, H.~Sun, and M.~Skoglund.
\newblock The capacity of private information retrieval with eavesdroppers.
\newblock {\em IEEE Trans. Inf. Theory}, 65(5):3198--3214, December 2018.

\bibitem{kadhe_singleserver_pir}
A.~Heidarzadeh, S.~Kadhe, S.~El Rouayheb, and A.~Sprintson.
\newblock Single-server multi-message individually-private information retrieval with side information.
\newblock In {\em IEEE ISIT}, July 2019.

\bibitem{ulukusPIRLC}
S.~Ulukus, S.~Avestimehr, M.~Gastpar, S.~A. Jafar, R.~Tandon, and C.~Tian.
\newblock Private retrieval, computing, and learning: Recent progress and future challenges.
\newblock {\em IEEE J. Sel. Areas Commun.}, 40(3):729--748, March 2022.

\bibitem{spir_first}
Y.~Gertner, Y.~Ishai, E.~Kushilevitz, and T.~Malkin.
\newblock Protecting data privacy in private information retrieval schemes.
\newblock {\em J. Computer System Sciences}, 60(3):592--629, 2000.

\bibitem{c_spir}
H.~Sun and S.~A. Jafar.
\newblock The capacity of symmetric private information retrieval.
\newblock {\em IEEE Trans. Inf. Theory}, 65(1):322--329, June 2018.

\bibitem{skoglund_mds_spir}
Q.~Wang and M.~Skoglund.
\newblock Symmetric private information retrieval from {MDS} coded distributed storage with non-colluding and colluding servers.
\newblock {\em IEEE Trans. Inf. Theory}, 65(8):5160--5175, March 2019.

\bibitem{sun_spir_mds_mismatched}
Q.~Wang, H.~Sun, and M.~Skoglund.
\newblock Symmetric private information retrieval with mismatched coded messages and randomness.
\newblock In {\em IEEE ISIT}, July 2019.

\bibitem{pir_spir_adversaries}
Q.~Wang and M.~Skoglund.
\newblock On {PIR} and symmetric {PIR} from colluding databases with adversaries and eavesdroppers.
\newblock {\em IEEE Trans. Inf. Theory}, 65(5):3183--3197, October 2019.

\bibitem{wang_mmspir}
Z.~Wang, K.~Banawan, and S.~Ulukus.
\newblock Private set intersection: A multi-message symmetric private information retrieval perspective.
\newblock {\em IEEE Trans. Inf. Theory}, 68(3):2001--2019, November 2021.

\bibitem{zhusheng_spir_pir}
Z.~Wang and S.~Ulukus.
\newblock Symmetric private information retrieval at the private information retrieval rate.
\newblock {\em IEEE J. Sel. Areas Inf. Theory}, 3(2):350--361, 2022.

\bibitem{ali_dapac}
A.~M. Jafarpisheh, M.~Mirmohseni, and M.~A. Maddah-Ali.
\newblock Distributed attribute-based private access control.
\newblock In {\em IEEE ISIT}, June 2022.

\bibitem{meel_hetdapac}
S.~Meel and S.~Ulukus.
\newblock {HetDAPAC}: Distributed attribute-based private access control with heterogeneous attributes.
\newblock In {\em IEEE ISIT}, July 2024.

\bibitem{graphbased_pir}
N.~Raviv, I.~Tamo, and E.~Yaakobi.
\newblock Private information retrieval in graph-based replication systems.
\newblock {\em IEEE Trans. Inf. Theory}, 66(6):3590--3602, November 2019.

\bibitem{BU19}
K.~Banawan and S.~Ulukus.
\newblock Private information retrieval from non-replicated databases.
\newblock In {\em IEEE ISIT}, July 2019.

\bibitem{SGT23}
B.~Sadeh, Y.~Gu, and I.~Tamo.
\newblock Bounds on the capacity of private information retrieval over graphs.
\newblock {\em IEEE Trans. Inf. Forensics Security}, 18:261--273, November 2023.

\bibitem{meel_multi_pir}
S.~Meel, X.~Kong, T.~J. Maranzatto, I.~Tamo, and S.~Ulukus.
\newblock Private information retrieval on multigraph-based replicated storage.
\newblock In {\em IEEE ISIT}, June 2025.

\bibitem{meel2025symmetric}
S.~Meel and S.~Ulukus.
\newblock Symmetric private information retrieval ({SPIR}) on graph-based replicated systems.
\newblock In {\em IEEE Globecom}, December 2025.

\bibitem{our_journal2025}
X.~Kong, S.~Meel, T.~J. Maranzatto, I.~Tamo, and S.~Ulukus.
\newblock New capacity bounds for {PIR} on graph and multigraph-based replicated storage, 2025.
\newblock {Available} online at arXiv:2504.20888.

\end{thebibliography}
\end{document}